\documentclass[onecolumn,showkeys,preprintnumbers,amsmath,amssymb]{revtex4-2}

\usepackage{graphicx}
\usepackage{dcolumn}
\usepackage{bm}
\usepackage[utf8]{inputenc}
\usepackage{amsmath}
\usepackage{amssymb}
\usepackage{booktabs}
\usepackage{color}
\usepackage{units}	
\usepackage{subcaption}
\usepackage{float}
\usepackage{svg}
\usepackage{changepage}
\usepackage{comment}

\begin{document}


\title{Anisotropic Bianchi-I cosmological model in non-conservative unimodular gravity.}
\author{Marcelo H. Alvarenga}
 \altaffiliation{}
 \email{marcelo.alvarenga@edu.ufes.br}
\affiliation{Núcleo Cosmo-ufes \& Departamento de Física, UFES, Vitória, ES, Brazil
}%
\author{Júlio C. Fabris}
 \altaffiliation{}
 \email{julio.fabris@cosmo-ufes.org}
\affiliation{Núcleo Cosmo-ufes \& Departamento de Física, UFES, Vitória, ES, Brazil
\\
National Research Nuclear University MEPhI, Kashirskoe sh. 31, Moscow 115409, Russia
}%



\date{\today}

\begin{abstract}
In this article, we propose an anisotropic Bianchi-I type cosmological model in non-conservative Unimodular Gravity ($\mathrm{NUG}$). We show that simply using the Bianchi-I type metric does not resolve a striking characteristic of the field equations in $\mathrm{NUG}$: their underdetermination. This fact led us to implement extra conditions on the combination $\left(\rho+p\right)$ and, consequently, obtain a consistent background cosmological analysis. In the vacuum case, we obtain an analogy between the Kasner solutions and the equations in $\mathrm{NUG}$. We also propose a new analysis of a non-homogeneous equation of state, the combination $\left(\rho+p\right)=l$. We identify that the cosmological dynamics are strictly dependent on the value of the constant $l$. The physically interesting case is at the value $l<0$, which seems to indicate a super-accelerated, ghost-like universe. This case still requires a more detailed analysis, for example, from a thermodynamic point of view, keeping in mind that $\left(\rho+p\right)$ may be interpreted as enthalpy of the system. For the cases $\left(\rho+p\right)\propto a^{-3}$ and $\left(\rho+p\right) \propto a^{-4}$, we obtain a description consistent with the anisotropic cosmological model described by $\mathrm{GR}$. In all cases analyzed, a small value for the anisotropic parameter $\Omega_{A}$ (on the order of $10^{-2}$) is required in order to have agreement, for example, with the age of the universe to be approximately $12-14\, \mathrm{Gyr}$, agreeing with the age of globular clusters. As the universe expands an isotropization is verified, with the anistropies going to zero asymptotically, similarly with what happens in an anistropic cosmological model based on $\mathrm{GR}$.

\end{abstract}

\maketitle

\section{Introduction}
After just over a century of its theoretical development, General Relativity $\mathrm{GR}$ \cite{Einstein:1916vd} remains the subject of considerable debate in the literature. Despite its success in explaining several phenomena (such as the theoretical prediction of black holes and gravitational waves, both confirmed by observations), the theory has difficulties in explaining other
phenomena. The Standard Cosmological Scenario, for example, which is based on GR, needs the introduction of a
dark sector which remains without any direct detection. Furthermore, $\mathrm{GR}$ predicts the existence of singularities either
inside black holes and in the early universe. These theoretical predictions are a great challenge for the theory. 
On the other side, applying GR theory to cosmology, together with the cosmological principle, which proposes
spatial homogeneity and isotropy on large scales of the Universe, leads to a successful theoretical model in describing
the Universe as a whole: the Standard Cosmological Model $\Lambda \mathrm{CDM}$ \cite{Peebles:2002gy, Zeldovich:1968ehl, Weinberg}. It simply explains the current phase
of accelerated expansion of the universe and give the best fits observational data. However, as already mentioned,
it must introduce two exotic components of unknown origin into its description of the dark sector: dark matter and
dark energy \cite{Peebles:2002gy, Weinberg}. Regarding the data, recent observations indicate unexpected features in the microwave sky, such as the absence of variance and correlation at the largest angular scales and the alignment of multipolar moments \cite{Bull:2015stt, Schwarz:2015cma, Perivolaropoulos:2021jda}, and this seems to indicate a violation of statistical isotropy at large scales \cite{Schwarz:2015cma, Aluri:2022hzs}. Indeed, even though observations of the cosmic microwave background $(\mathrm{CMB})$ \cite{Planck:2018nkj, WMAP:2012fli} is in agreement with $\Lambda \mathrm{CDM}$ model based on the cosmological principle, it seems to exhibited several anisotropic features on large angular scales, and these are the subject of studies in the literature,
see \cite{Bennett:2010jb, Planck:2019evm}. This motivates us to explore theoretical models that exhibit a violation of the cosmological principle while maintaining spatial homogeneity and introducing new forms of anisotropy. Numerous studies have been carried out in this regard on anisotropic models  \cite{1968ApJ...153..661J, Mishra:2017ycy, Galeev:2021xit, Parnovsky:2022cyb, Parnovsky:2023hzo, Sharma:2024sjk}. An example in the context of dark energy can be seen in \cite{Verma:2024lex, Yadav:2011bj} and references therein.

On the other hand, the dark energy responsible for the current accelerated expansion of the Universe is driven (modeled) by the Cosmological Constant $(\mathrm{CC})$ $\Lambda$ \cite{Zeldovich:1968ehl, Copeland:2006wr,Bamba:2012cp} where, at this point, one of the fundamental problems of theoretical physics in current times is concentrated: the $\mathrm{CC}$ problem \cite{Bull:2015stt, Martin:2012bt, RevModPhys.61.1}. The big question surrounding this problem can be briefly posed as whether we have an interpretation of the $\mathrm{CC}$ as a fluid related to the energy density of the quantum vacuum. This leads to a huge discrepancy between the theoretical value (coming from quantum field theory about the quantum vacuum value) and its observational value (coming from the standard cosmological model $\Lambda \mathrm{CDM}$), estimated to be up to $120$ orders of magnitude \cite{RevModPhys.61.1}. Although, this number can be reduced (leading to $60$, $50$ orders of magnitude) in more detailed calculations introducing, for example, concepts of supersymmetries \cite{Martin:2012bt}, the discrepancy between theory and observation remains huge. Furthermore, its nature and behavior still cause discomfort for many expert theorists. Questions like where did the $\mathrm{CC}$ come from? And is it truly a constant throughout all phases of the Universe? These questions still lack a concrete and definitive answer. Although we will not present an answer to these questions in this work (because this is still under constant construction and follows several different directions), we will explore an alternative model to $\mathrm{GR}$ in the cosmological context known today as Unimodular Gravity that allows us a different conception of $\mathrm{CC}$.

Shortly after $\mathrm{GR's}$ theoretical development, several issues were debated, and new alternative theories were created in the quest for a better understanding of gravitational phenomena. And the author of GR himself, Albert Einstein, initiated what is now considered one of the simplest alternative theories in by restricting the determinant of the metric tensor $\sqrt{-g}=1$, obtaining a considerable simplification in the search for solutions to the gravitational field equations \cite{Einstein:1919gv}. Later, this restriction re-elaborated through the principle of least action in which the condition on the determinant of the metric tensor would appear as a constraint in the system \cite{Anderson:1971pn, 1991JMP....32.1337N}. This approach was named Unimodular Theory or simply Unimodular Gravity ($\mathrm{UG}$). One of the main features of this theory is its trace-free field equations. A direct consequence of this is the absence of information about, for example, geometric quantities such as the Ricci scalar. Assuming conservation of the energy-momentum tensor as an extra condition in the solution of the complete set of equations in the cosmological context yields exactly $\mathrm{GR}+\Lambda$ ; however, here $\Lambda$ is a simple constant of integration associated with $\mathrm{CC}$. This has sometimes been considered as a possible solution to the $\mathrm{CC}$ problem, see \cite{Ellis:2010uc, Ellis:2013uxa, RevModPhys.61.1}. We also highlight that the condition trace-free of the equations makes it sensitive only to the combination $\left(\rho+p\right)$ which, from a thermodynamic point of view, represents the enthalpy \cite{Fabris_2022}. Furthermore, unlike in $\mathrm{GR}$, the Bianchi Identities ($\mathrm{BI}$) in $\mathrm{UG}$ do not lead directly to conservation of the energy-momentum tensor, which makes it possible to explore it as a non-conservative theory, see \cite{1993JMP....34.2465T, Fabris:2023vop}. Although this can be somewhat mapped into $\mathrm{GR}$ as a $\Lambda(t)$, this possibility has been gaining prominence in the literature, especially in the cosmological context, as a discussion about a dynamic $\mathrm{CC}$ $\Lambda\left(t\right)$ in $\mathrm{UG}$ \cite{Alvarenga:2024yqa} or even a possible solution to the tension problem in $H_{0}$ \cite{Perez:2020cwa, Plaza:2025nip}. However, when the conservation of the energy-momentum tensor is not separately imposed, an underdetermined system of equations is obtained and its solution will only be possible if an extra condition is imposed see \cite{Alvarenga:2024yqa, Garcia-Aspeitia:2019yod,Fabris_2022, Fabris:2023vop}. In this case, a parallel to the the $\mathrm{GR}$ approach to a dynamic cosmological
term can be traced. Given the many open questions still related to both $\mathrm{GR}$ and $\mathrm{UG}$, we would like to propose an analysis of anisotropic Bianchi I-type models in non-conservative unimodular theory. It is worth highlighting that the term 'non-conservative' means generalizing the usual minimal conservation laws, introducing a new connection between geometry and matter fields. It should be kept in mind that this term does not mean abandoning the general conservation laws in the strict sense. We will explore the features arising from trace-free equations and non-conservation. Under these conditions, can we achieve a system that is not underdetermined? How does the universe evolve for specific cases of the combination $\left(\rho+p\right)$? Can we find a viable model description for our universe? These are some of the questions we would like to discuss in this paper.

The paper is structured as follows: Section (\ref{sec:2}) contains the basic equations describing an anisotropic cosmological model in non-conservative Unimodular Gravity. Once the equations that will dictate the cosmological dynamics have been described, we move on to analyzing some specific cases for the combination $\left(\rho+p\right)$. The analysis for the vacuum case is found in Section (\ref{sec:3}). Section (\ref{sec:4}) dedicates the analysis to the case $\left(\rho+p\right)=\mathrm{cte}$, that is, when we have a constant enthalpy. Section (\ref{sec:5}) is dedicated to analyzing the case where we have a fluid with typical dust behavior $p=0$, that is, $\rho_{m}=\left(\frac{a}{a_{0}}\right)^{-3}$. Section (\ref{sec:6}) is dedicated to analyzing the case where we have a fluid with typical radiation behavior, that is, $\rho_{r}=\left(\frac{a}{a_{0}}\right)^{-4}$. And finally, Section (\ref{sec:7}) is dedicated to the final considerations.

\section{Basic equations.}\label{sec:2}

The simplest anisotropic cosmological model, which nevertheless fully
describes anisotropic effects, is the so-called homogeneous Bianchi
type I model. In this model, the spatial sections are flat, but
the rate of expansion or contraction depends on the direction. The
gravitational field for a spatially homogeneous and anisotropic is
described by metric as
\begin{align}
ds^{2} & =dt^{2}-A\left(t\right)^{2}dx^{2}-B\left(t\right)^{2}dy^{2}-C\left(t\right)^{2}dz^{2},\label{eq:1}
\end{align}
where $A$, $B$ e $C$ are functions of cosmic time $t$.

The spatial volume ($V$) and the average scale factor ($a$) for
Bianchi
type I spacetime are given by
\begin{align}
V & =ABC\label{eq:2}\\
a & =\left(ABC\right)^{\frac{1}{3}}=V^{\frac{1}{3}}.\label{eq:3}
\end{align}
The Hubble expansion rates for Bianchi
type I is given by
\begin{align}
H=\frac{\dot{a}}{a} & =\frac{1}{3}\left(\frac{\dot{A}}{A}+\frac{\dot{B}}{B}+\frac{\dot{C}}{C}\right)=\frac{1}{3}\left(H_{1}+H_{2}+H_{3}\right),\label{eq:4}
\end{align}
where an overdot represents partial differentiation with respect to
cosmic time $t$ and we have defined $H_{1}\equiv\frac{\dot{A}}{A}$,
$H_{2}\equiv\frac{\dot{B}}{B}$ and $H_{3}\equiv\frac{\dot{C}}{C}$
as the directional Hubble parameter along the $x$, $y$ and $z$
axes, respectively.

The non-zero Christoffel symbols $\left(\Gamma_{\mu\nu}^{\alpha}\right)$
are
\begin{align}
\Gamma_{11}^{0}=A\dot{A};\,\, & \Gamma_{22}^{0}=B\dot{B};\,\,\Gamma_{33}^{0}=C\dot{C}\nonumber \\
\Gamma_{10}^{1}=\frac{\dot{A}}{A};\,\, & \Gamma_{20}^{2}=\frac{\dot{B}}{B};\,\,\Gamma_{30}^{3}=\frac{\dot{C}}{C}.\label{eq:5}
\end{align}
We can use the components of the Christoffel symbols, expression (\ref{eq:5}),
and obtain the non-zero components of the Ricci tensor $\left(R_{\mu\nu}\right)$,
given by
\begin{align}
R_{00} & =-\left[\frac{\ddot{A}}{A}+\frac{\ddot{B}}{B}+\frac{\ddot{C}}{C}\right]\label{eq:6}\\
R_{11} & =A^{2}\left[\frac{\ddot{A}}{A}+\frac{\dot{A}}{A}\frac{\dot{B}}{B}+\frac{\dot{A}}{A}\frac{\dot{C}}{C}\right]\label{eq:7}\\
R_{22} & =B^{2}\left[\frac{\ddot{B}}{B}+\frac{\dot{A}}{A}\frac{\dot{B}}{B}+\frac{\dot{B}}{B}\frac{\dot{C}}{C}\right]\label{eq:8}\\
R_{33} & =C^{2}\left[\frac{\ddot{C}}{C}+\frac{\dot{A}}{A}\frac{\dot{C}}{C}+\frac{\dot{B}}{B}\frac{\dot{C}}{C}\right].\label{eq:9}
\end{align}
The scalar curvature $\left(R\right)$ is given by
\begin{align}
R & =-2\left[\frac{\ddot{A}}{A}+\frac{\ddot{B}}{B}+\frac{\ddot{C}}{C}+\frac{\dot{A}}{A}\frac{\dot{B}}{B}+\frac{\dot{A}}{A}\frac{\dot{C}}{C}+\frac{\dot{B}}{B}\frac{\dot{C}}{C}\right].\label{eq:10}
\end{align}

The field equations in Non-Conservative Unimodular Gravity ($\mathrm{NUG}$) is given
by
\begin{align}
R_{\mu\nu}-\frac{1}{4}g_{\mu\nu}R & =8\pi G\left(T_{\mu\nu}-\frac{1}{4}g_{\mu\nu}T\right),\label{eq:11}\\
\frac{R^{;\nu}}{4} & =8\pi G\left(T_{;\mu}^{\mu\nu}-\frac{1}{4}T^{;\nu}\right).\label{eq:12}
\end{align}
The expression (\ref{eq:11}) are trace-free equations, while the
expression (\ref{eq:12}) is the result when we apply the Bianchi
identities in the expression (\ref{eq:11}). The name "non-conservative"
means that we will not impose the usual conservation of the energy-momentum
tensor in the cosmological analysis in the unimodular theory. We will
keep the expression (\ref{eq:12}) in its form as it is in the description
of an anisotropic cosmological model.

We will initially consider the material content of the universe as
consisting of perfect fluid, whose expression is given by the energy-momentum
tensor as
\begin{align}
T_{\mu\nu} & =\left(\rho+p\right)u_{\mu}u_{\nu}-g_{\mu\nu}p,
\end{align}
where $u_{\mu}$ is the $4$-velocity of the perfect fluid in the
coordinate system of the expression (\ref{eq:1}), satisfying the
condition $u^{\mu}u_{\mu}=1$. Therefore, the non-zero components
of the energy-momentum tensor and its trace will be given, from (\ref{eq:1}),
as
\begin{align}
T_{00}=\rho;\,\, & T_{11}=A^{2}p;\,\,T_{22}=B^{2}p;\\
T_{33} & =C^{2}p;\,\,T=\rho-3p.
\end{align}

The solutions of the field equations (\ref{eq:11}) give us the following
equations
\begin{align}
2\left\{ \frac{\ddot{A}}{A}+\frac{\ddot{B}}{B}+\frac{\ddot{C}}{C}-\frac{\dot{A}}{A}\frac{\dot{B}}{B}-\frac{\dot{A}}{A}\frac{\dot{C}}{C}-\frac{\dot{B}}{B}\frac{\dot{C}}{C}\right\}  & =-24\pi G\left(\rho+p\right)\label{eq:16}\\
2\left\{ \frac{\ddot{A}}{A}-\frac{\ddot{B}}{B}-\frac{\ddot{C}}{C}+\frac{\dot{A}}{A}\frac{\dot{B}}{B}+\frac{\dot{A}}{A}\frac{\dot{C}}{C}-\frac{\dot{B}}{B}\frac{\dot{C}}{C}\right\}  & =8\pi G\left(\rho+p\right)\label{eq:17}\\
2\left\{ \frac{\ddot{B}}{B}-\frac{\ddot{A}}{A}-\frac{\ddot{C}}{C}+\frac{\dot{A}}{A}\frac{\dot{B}}{B}-\frac{\dot{A}}{A}\frac{\dot{C}}{C}+\frac{\dot{B}}{B}\frac{\dot{C}}{C}\right\}  & =8\pi G\left(\rho+p\right)\label{eq:18}\\
2\left\{ \frac{\ddot{C}}{C}-\frac{\ddot{A}}{A}-\frac{\ddot{B}}{B}-\frac{\dot{A}}{A}\frac{\dot{B}}{B}+\frac{\dot{A}}{A}\frac{\dot{C}}{C}+\frac{\dot{B}}{B}\frac{\dot{C}}{C}\right\}  & =8\pi G\left(\rho+p\right).\label{eq:19}
\end{align}

In a direct analysis of expressions (\ref{eq:17})-(\ref{eq:19})
their sum leads to the expression (\ref{eq:16}), as they indeed must
be due to the trace-free nature of the equations in $\mathrm{UG}$. On the other
hand, the solution of "non-conservation" equation, expression (\ref{eq:12})
is given by
\begin{align}
\frac{\dddot{A}}{A}+\frac{\dddot{B}}{B}+\frac{\dddot{C}}{C}+ & \left\{ \frac{d}{dt}\left[\frac{\dot{B}}{B}+\frac{\dot{C}}{C}\right]-\frac{\ddot{A}}{A}\right\} \frac{\dot{A}}{A}+\left\{ \frac{d}{dt}\left[\frac{\dot{A}}{A}+\frac{\dot{C}}{C}\right]-\frac{\ddot{B}}{B}\right\} \frac{\dot{B}}{B}+\nonumber \\
+\left\{ \frac{d}{dt}\left[\frac{\dot{A}}{A}+\frac{\dot{B}}{B}\right]-\frac{\ddot{C}}{C}\right\} \frac{\dot{C}}{C} & =-16\pi G\left\{ \frac{3}{4}\left(\dot{\rho}+\dot{p}\right)+\left[\frac{\dot{A}}{A}+\frac{\dot{B}}{B}+\frac{\dot{C}}{C}\right]\left(\rho+p\right)\right\} .\label{eq:20}
\end{align}

To analyze the solutions, equations (\ref{eq:16})-(\ref{eq:20})
we will introduce the following changes in variables
\begin{align}
A\equiv a\left(t\right)\exp\left[\alpha\left(t\right)\right];\,\,\,\,B\equiv a\left(t\right)\exp\left[\beta\left(t\right)\right] & ;\,\,\,\,C\equiv a\left(t\right)\exp\left[\gamma\left(t\right)\right],\label{eq:21}
\end{align}
where $a\left(t\right)$ is the average scale factor and $\alpha\left(t\right)$,
$\beta\left(t\right)$, $\gamma\left(t\right)$ are anisotropic functions.
It is easy to see that the condition (\ref{eq:3}) causes the anisotropic
functions to satisfy the following relation
\begin{align}
\alpha\left(t\right)+\beta\left(t\right)+\gamma\left(t\right) & =0.\label{eq:22}
\end{align}
Under this new change, the spacial volume will be given by
\begin{align*}
V & =ABC=a\left(t\right)^{3}.
\end{align*}
Therefore the metric takes the form
\begin{align}
ds^{2} & =dt^{2}-a^{2}\left(t\right)\left\{\exp\left[2\alpha\left(t\right)\right]dx^{2}+\exp\left[2\beta\left(t\right)\right]dy^{2}-\exp\left[2\gamma\left(t\right)\right]dz^{2}\right\},\label{eq:metric}
\end{align}
the metric in the form (\ref{eq:metric}) has been introduced in the
paper \cite{1968ApJ...153..661J}.

We can rewrite the equations (\ref{eq:16})-(\ref{eq:20}) using these
new variables and consequently the relation (\ref{eq:22}), obtaining
\begin{align}
\frac{\ddot{a}}{a}-\frac{\dot{a}^{2}}{a^{2}}+\left[\dot{\alpha}^{2}+\dot{\alpha}\dot{\beta}+\dot{\beta}^{2}\right] & =-4\pi G\left(\rho+p\right)\label{eq:23}\\
-\frac{\ddot{a}}{a}+\frac{\dot{a}^{2}}{a^{2}}+6\dot{\alpha}\frac{\dot{a}}{a}+2\ddot{\alpha}-\left[\dot{\alpha}^{2}+\dot{\alpha}\dot{\beta}+\dot{\beta}^{2}\right] & =4\pi G\left(\rho+p\right)\label{eq:24}\\
-\frac{\ddot{a}}{a}+\frac{\dot{a}^{2}}{a^{2}}+6\dot{\beta}\frac{\dot{a}}{a}+2\ddot{\beta}-\left[\dot{\alpha}^{2}+\dot{\alpha}\dot{\beta}+\dot{\beta}^{2}\right] & =4\pi G\left(\rho+p\right)\label{eq:25}\\
-\frac{\ddot{a}}{a}+\frac{\dot{a}^{2}}{a^{2}}-6\left\{ \dot{\alpha}+\dot{\beta}\right\} \frac{\dot{a}}{a}-2\left(\ddot{\alpha}+\ddot{\beta}\right)-\left[\dot{\alpha}^{2}+\dot{\alpha}\dot{\beta}+\dot{\beta}^{2}\right] & =4\pi G\left(\rho+p\right)\label{eq:26}\\
3\frac{\dddot{a}}{a}+3\frac{\ddot{a}\dot{a}}{a^{2}}-6\frac{\dot{a}^{3}}{a^{3}}+2\left(\ddot{\alpha}\dot{\alpha}+\ddot{\beta}\dot{\beta}\right)+\ddot{\alpha}\dot{\beta}+\ddot{\beta}\dot{\alpha} & =\nonumber \\
=-16\pi G\left\{ \frac{3}{4}\left(\dot{\rho}+\dot{p}\right)+3\frac{\dot{a}}{a}\left(\rho+p\right)\right\} .\label{eq:27}
\end{align}

Finally, introducing two new independent anisotropic functions that
are defined in the $\left(x,y\right)$ plane and perpendicular to
it given by
\begin{align}
\eta\equiv\alpha+\beta; & \,\,\,\,\chi\equiv\alpha-\beta.\label{eq:28}
\end{align}
 Using (\ref{eq:28}) into the equations (\ref{eq:23})-(\ref{eq:27})
and combining the expressions (\ref{eq:24})-(\ref{eq:26}) results
in the complete set equations that will describe the cosmological
dynamics of the anisotropic model, they are
\begin{align}
\frac{\ddot{a}}{a}-\frac{\dot{a}^{2}}{a^{2}}+\frac{1}{4}\left[3\dot{\eta}^{2}+\dot{\chi}^{2}\right] & =-4\pi G\left(\rho+p\right)\label{eq:29}\\
\ddot{\eta}+3\left(\frac{\dot{a}}{a}\right)\dot{\eta} & =0\label{eq:30}\\
\ddot{\chi}+3\left(\frac{\dot{a}}{a}\right)\dot{\chi} & =0\label{eq:31}\\
3\frac{\dddot{a}}{a}+3\frac{\ddot{a}\dot{a}}{a^{2}}-6\frac{\dot{a}^{3}}{a^{3}}+\frac{3}{2}\ddot{\eta}\dot{\eta}+\frac{1}{2}\ddot{\chi}\dot{\chi} & =-16\pi G\left\{ \frac{3}{4}\left(\dot{\rho}+\dot{p}\right)+3\frac{\dot{a}}{a}\left(\rho+p\right)\right\} .\label{eq:32}
\end{align}

The isotropic limit occurs when $A=B=C=a\left(t\right)$, for this
$\alpha$, $\beta$, $\gamma$ go to zero and, consequently, we then
obtain the same equations studied in the isotropic case.

The expressions (\ref{eq:30}) and (\ref{eq:31}) can be easily solved
analytically, obtaining
\begin{align}
\dot{\eta}a^{3}=X_{1}a_{0}^{3}; & \,\,\,\,\dot{\chi}a^{3}=X_{2}a_{0}^{3}.\label{eq:33}
\end{align}
The expressions in (\ref{eq:33}) directly imply
\begin{align}
\eta=X_{1}\int\left(\frac{a}{a_{0}}\right)^{-3}dt; & \,\,\,\,\chi=X_{2}\int\left(\frac{a}{a_{0}}\right)^{-3}dt,\label{eq:34}
\end{align}
where $X_{1}$, $X_{2}$ are constants of integration and $a_{0}$
denotes the value of $a$ at some fixed proper time, i.e, $a\left(t=t_{0}\right)=a_{0}$.
We can use the expressions in (\ref{eq:34}) in the equations (\ref{eq:29})
and (\ref{eq:32}), obtaining
\begin{align}
\frac{\ddot{a}}{a}-\frac{\dot{a}^{2}}{a^{2}}+\frac{1}{4}\left[3X_{1}^{2}+X_{2}^{2}\right]\left(\frac{a}{a_{0}}\right)^{-6} & =-4\pi G\left(\rho+p\right)\label{eq:35}\\
3\frac{\dddot{a}}{a}+3\frac{\ddot{a}\dot{a}}{a^{2}}-6\frac{\dot{a}^{3}}{a^{3}}-\frac{3}{2}\left\{ 3X_{1}^{2}+X_{2}^{2}\right\} \left(\frac{a}{a_{0}}\right)^{-7}\left(\frac{\dot{a}}{a_{0}}\right) & =-16\pi G\left\{ \frac{3}{4}\left(\dot{\rho}+\dot{p}\right)+3\frac{\dot{a}}{a}\left(\rho+p\right)\right\} .\label{eq:36}
\end{align}
When using the equation (\ref{eq:35}) inside the expression (\ref{eq:36})
we still obtain an underdetermined system $0=0$.

In the next sections we will analyze some solutions to the equation
(\ref{eq:35}) and (\ref{eq:36}). 

\section{Case $\left(\rho+p\right)=0$.}\label{sec:3}

When we have $\left(\rho+p \right)=0$, the equation (\ref{eq:35}) and (\ref{eq:36})
is as follows
\begin{align}
\frac{\ddot{a}}{a}-\frac{\dot{a}^{2}}{a^{2}}+\frac{1}{4}\left[3X_{1}^{2}+X_{2}^{2}\right]\left(\frac{a}{a_{0}}\right)^{-6} & =0\label{eq:37}\\
\frac{\dddot{a}}{a}+\frac{\ddot{a}\dot{a}}{a^{2}}-2\frac{\dot{a}^{3}}{a^{3}}-\frac{1}{2}\left[3X_{1}^{2}+X_{2}^{2}\right]\left(\frac{a}{a_{0}}\right)^{-7}\left(\frac{\dot{a}}{a_{0}}\right) & =0\label{eq:38n}
\end{align}
The expression (\ref{eq:38n}) can be rewritten as follows
\begin{align}
\frac{d}{dt}\left[\frac{\ddot{a}}{a}+\frac{\dot{a}^{2}}{a^{2}}+\frac{\left[3X_{1}^{2}+X_{2}^{2}\right]}{12}\left(\frac{a}{a_{0}}\right)^{-6}\right] & =0.\label{eq:39n}
\end{align}
The expression (\ref{eq:39n}) implies that
\begin{align}
\frac{\ddot{a}}{a}+\frac{\dot{a}^{2}}{a^{2}}+\frac{\left[3X_{1}^{2}+X_{2}^{2}\right]}{12}\left(\frac{a}{a_{0}}\right)^{-6} & =\frac{2}{3}\Lambda_{U},\label{eq:40n}
\end{align}
where $\Lambda_{U}$ is a constant that we designate as the cosmological
constant. Using the expressions (\ref{eq:37}) and (\ref{eq:40n}) the combination becomes
\begin{align}
\frac{\dot{a}^{2}}{a^{2}} & =\frac{\Lambda_{U}}{3}+\frac{\left[3X_{1}^{2}+X_{2}^{2}\right]}{12}\left(\frac{a}{a_{0}}\right)^{-6}.\label{eq:41n}
\end{align}
Now, we can manipulate the expression (\ref{eq:41n}) and write it in its differential form
\begin{align}
\frac{a^{2}da}{\sqrt{a^{6}+\frac{\Omega_{A}}{\Omega_{U}}a_{0}^{6}}}	=\sqrt{\Omega_{U}}H_{0}dt,\label{eq:43n1}
\end{align}
where we have defined 
$\Omega_{A}\equiv\frac{\left[3X_{1}^{2}+X_{2}^{2}\right]}{12H_{0}^{2}}$ e $\Omega_{U}\equiv\frac{\Lambda_{U}}{3H_{0}^{2}}$.
The expression (\ref{eq:43n1}) can be integraded and we get as solution
\begin{align}
\frac{a\left(t\right)}{a_{0}}	=\left(\frac{\Omega_{A}}{\Omega_{U}}\right)^{\frac{1}{6}}\sinh^{\frac{1}{3}}\left(3\sqrt{\Omega_{U}}H_{0}t\right).\label{eq:at}
\end{align}

Now, if we compare the anisotropic cosmological model described by
$\mathrm{NUG}$ with the $\mathrm{GR}+\Lambda$ case, there is no difference
when we assume the combination $\left(\rho+p\right)=0$, although
the cosmological constant in $\mathrm{NUG}$ originates from an integration
constant. We can clearly see this if we look at the field equations
that describe cosmological dynamics according to the anisotropic model
described by $\mathrm{GR}$ in presence of matter defined by an equation of state $p = \omega\rho$, given by
\begin{align}
3\left(\frac{\dot{a}}{a}\right)^{2}-\frac{\left[3\dot{\eta}^{2}+\dot{\chi}^{2}\right]}{4} & =8\pi G\rho_{0}\left(\frac{a}{a_{0}}\right)^{-3\left(1+\omega\right)}\label{eq:57n}\\
\ddot{\eta}+3\left(\frac{\dot{a}}{a}\right)\dot{\eta} & =0\label{eq:58n}\\
\ddot{\chi}+3\left(\frac{\dot{a}}{a}\right)\dot{\chi} & =0,\label{eq:59n}
\end{align}
where $\omega$ is the constant equation of state parameter,
$a_{0}$ is the value $a$ at a fixed time and the functions $\left(\eta,\chi\right)$
are the anisotropic functions identical to those defined in (\ref{eq:28}). We can express the solutions (\ref{eq:57n})-(\ref{eq:59n}) by specifying the equation of state parameter $\omega$, for example, in the case $\omega=0$ yields a universe model containing an anisotropic term and a fluid like dust permeating spacetime according to $\mathrm{GR}$.
We are also using the conservation equation of the energy-momentum
tensor $\left(\nabla_{\mu}T_{\nu}^{\mu}=0\right)$ which is given
by
\begin{align}
\dot{\rho}+\frac{\dot{V}}{V}\left(\rho+p\right) & =0,
\end{align}
where $V$ is spacial volume given by expression (\ref{eq:2}) which
consequently, using the change of variable given by (\ref{eq:21})
and relation (\ref{eq:22}) becomes: $V=a^{3}\left(t\right).$ Therefore,
the solution for energy density becomes
\begin{align}
\rho & =\rho_{0}\left(\frac{a}{a_{0}}\right)^{-3\left(1+\omega\right)},
\end{align}
which is properly represented on the right side of the equation (\ref{eq:57n}).
The solution to equations (\ref{eq:58n}) and (\ref{eq:59n}) have
already been obtained previously, they are given by expressions (\ref{eq:33})
and (\ref{eq:34}). Let us assume a fluid that behaves like the cosmological
constant that has equation of state $\rho_{\Lambda}=-p_{\Lambda}$,
being $\omega=-1$. Thus, the expression (\ref{eq:57n}) in terms
of ratio $a/a_{0}$ and using the solutions for $\dot{\eta}$ and
$\dot{\sigma}$ we obtain
\begin{align}
\left(\frac{\dot{a}}{a_{0}}\right)^{2} & =\frac{\Lambda}{3}\left(\frac{a}{a_{0}}\right)^{2}+\frac{\left\{ 3X_{1}^{2}+X_{2}^{2}\right\} }{12}\left(\frac{a}{a_{0}}\right)^{-4},\label{eq:62n}
\end{align}
where have defined $\rho_{0} \equiv \frac{\Lambda}{8\pi G}=cte$.

Note that the expression (\ref{eq:62n}) is the same as that abtained
when we assume the combination $\left(\rho+p\right)=0$ in the anisotropic
cosmological model in $\mathrm{NUG}$, the expression (\ref{eq:41n}). Thus demonstrating the equivalence between the two
approches. However, we emphasize that the cosmological constant in
$\mathrm{NUG}$ has its origins in an integration constant. 

\subsection{Kasner Solution in NUG.}\label{Kasner}

The vaccum solution $\left(\rho=p=0\right)$ was obtained in 1921 by Edward Kasner in the $\mathrm{GR}$ context with $\Lambda=0$ \cite{Kasner:1921zz}. This solution can be considered as a reference anisotropic cosmological model. The metric has the form \cite{Kasner:1921zz, Misner:1969hg, Belinskii_1971}
\begin{align}
    ds^{2}	=dt^{2}-t^{2p_{1}}dx^{2}-t^{2p_{2}}dy^{2}-t^{2p_{3}}dz^{2},\label{eq:1K}
\end{align}
where $p_{1}$, $p_{2}$ e $p_{3}$ are dimensionless indices, called Kasner indices. This indices satisfy two conditions
\begin{align}
    p_{1}+p_{2}+p_{3}=1;\,\,\,	p_{1}^{2}+p_{2}^{2}+p_{3}^{3}=1.\label{eq:2K}
\end{align}
The solution (\ref{eq:1K}) has a cosmological singularity in $t\rightarrow0$ corresponding to the birth of the Universe. As we will see below, the case $\left(\rho+p\right)=0$ for the anisotropic cosmological model in $\mathrm{NUG}$ leads to the Kasner solution. To do this, we must remember that the functions $A$, $B$, and $C$, expression (\ref{eq:21}), can be rewritten in terms of anisotropic functions according to
\begin{align}
    A= a\left(t\right)\exp\left(\frac{\eta+\chi}{2}\right);\,\,\,\,B= a\left(t\right)\exp\left(\frac{\eta-\chi}{2}\right) & ;\,\,\,\,C= a\left(t\right)\exp\left(-\eta\right),\label{eq:3K}
\end{align}
where we use the relations (\ref{eq:28}), (\ref{eq:22}) and the anisotropic functions $\eta$ and $\chi$ are given by the expression (\ref{eq:34}). Using the solution for the scaling factor of $a\left(t\right)$, expression (\ref{eq:at}), from (\ref{eq:34}) we find the anisotropic functions as
\begin{align}
    \eta=\frac{X_{1}}{3\sqrt{\Omega_{A}}H_{0}}\ln\left(\tanh\left(\frac{3}{2}\sqrt{\Omega_{U}}H_{0}t\right)\right)+C_{\eta};\,\,\,\,
\chi	=\frac{X_{2}}{3\sqrt{\Omega_{A}}H_{0}}\ln\left(\tanh\left(\frac{3}{2}\sqrt{\Omega_{U}}H_{0}t\right)\right)+C_{\chi},
\end{align}
where $C_{\eta}$ e $C_{\chi}$ are constants of integration. It is in the limit $t \rightarrow 0$ that Kasner solutions can be visualized, and in this limit the anisotropic functions $\eta$, $\chi$ and average scaling factor $a\left(t\right)$, expressed as (\ref{eq:at}) can be approximated.
\begin{align}
    \eta\approx\frac{X_{1}}{3\sqrt{\Omega_{A}}H_{0}}\ln\left(\frac{3}{2}\sqrt{\Omega_{U}}H_{0}t\right);\,\,\,\,
\chi	\approx\frac{X_{2}}{3\sqrt{\Omega_{A}}H_{0}}\ln\left(\frac{3}{2}\sqrt{\Omega_{U}}H_{0}t\right);\,\,\,\,a\left(t\right)\approx t^{1/3}
\end{align}
where we use the approximation $\tanh\left(at\right)\approx at$ and $\sinh\left(at\right)\approx at$ for $t \rightarrow 0$. Consequently, in this approximation, the functions $A$, $B$, and $C$ expression (\ref{eq:3K}) approximate power-law behavior in $t$. Therefore, we can relate the Kasner indices $p_{i}$ to the solutions of the anisotropic cosmological model in $\mathrm{NUG}$ (in the vacuum case), that is
\begin{align}
    p_{1}=\frac{1}{3}+\frac{\left(X_{1}+X_{2}\right)}{6\sqrt{\Omega_{A}}H_{0}};\,\,\,\,	p_{2}=\frac{1}{3}+\frac{\left(X_{1}-X_{2}\right)}{6\sqrt{\Omega_{A}}H_{0}};\,\,\,\,p_{3}=\frac{1}{3}-\frac{X_{1}}{3\sqrt{\Omega_{A}}H_{0}};
\end{align}
The first Kasner condition $\left(p_{1}+p_{2}+p_{3}=1\right)$ is automatically satisfied, since we define $a\left(t\right)$ as the average scaling factor and the condition $\alpha+\beta+\gamma=0$ in the parameterization ensures that anisotropies do not affect the average volume of the universe $V=ABC=a^{3}$. On the other hand, the second Kasner condition $\left(p_{1}^{2}+p_{2}^{2}+p_{3}^{2}=1\right)$ will select the possible values for $p_{i}$ (or $X_{1}$ and $X_{2}$) in such a way that
\begin{align}
    3X_{1}^{2}+X_{2}^{2}&=12H_{0}^{2}\Omega_{A}.\label{eq:K}
\end{align}
This is exactly how we defined our anisotropic parameter $\Omega_{A}$. Therefore, it is not arbitrary and is rooted in the very geometry of the Bianchi I type and in the compatibility of the field equations in $\mathrm{NUG}$ with the Kasner solutions of the $\mathrm{GR}$.

The anisotropic term that decrease with a factor $a^{-6}$, present
in the equations above, means that for very small values of the average
scale factor $a$ it dominates the constants, both for the case of
$\mathrm{GR}$ (constant $\Lambda$) and for the case of $\mathrm{NUG}$ (constant $\Lambda_{U}$)
leanding to the geometry in the contraction phase to the Kasner solution,
and also to BKL instabilities \cite{Kasner:1921zz,Belinskii_1971}. 

\subsection{Cosmological parameters.}

In this subsection we will analyze how the cosmological parameters
will behave. We begin by rewriting the expression (\ref{eq:41n}) in terms of Hubble parameter and in its version written in terms of the fractional energy
density parameters, i.e.
\begin{align}
H^{2} & =H_{0}^{2}\left[\Omega_{U}+\Omega_{A}\left(\frac{a}{a_{0}}\right)^{-6}\right],\label{eq:63hubble}
\end{align}
were $H_{0}$ is Hubble parameter today and we define $\Omega_{U}\equiv\frac{\Lambda_{U}}{3H_{0}^{2}}$,
$\Omega_{A}\equiv\frac{\left[3X_{1}^{2}+X_{2}^{2}\right]}{12H_{0}^{2}}$
as fractional energy density parameters. The isotropic limit occurs when $X_{1}$ and $X_2$ are identically zero, i.e., $\Omega_{A}=0$ in (\ref{eq:63hubble}) leading to the de Sitter scenario.

\begin{figure}[t]
    \begin{subfigure}[t]{0.48\textwidth}
     \includegraphics[width=\textwidth]{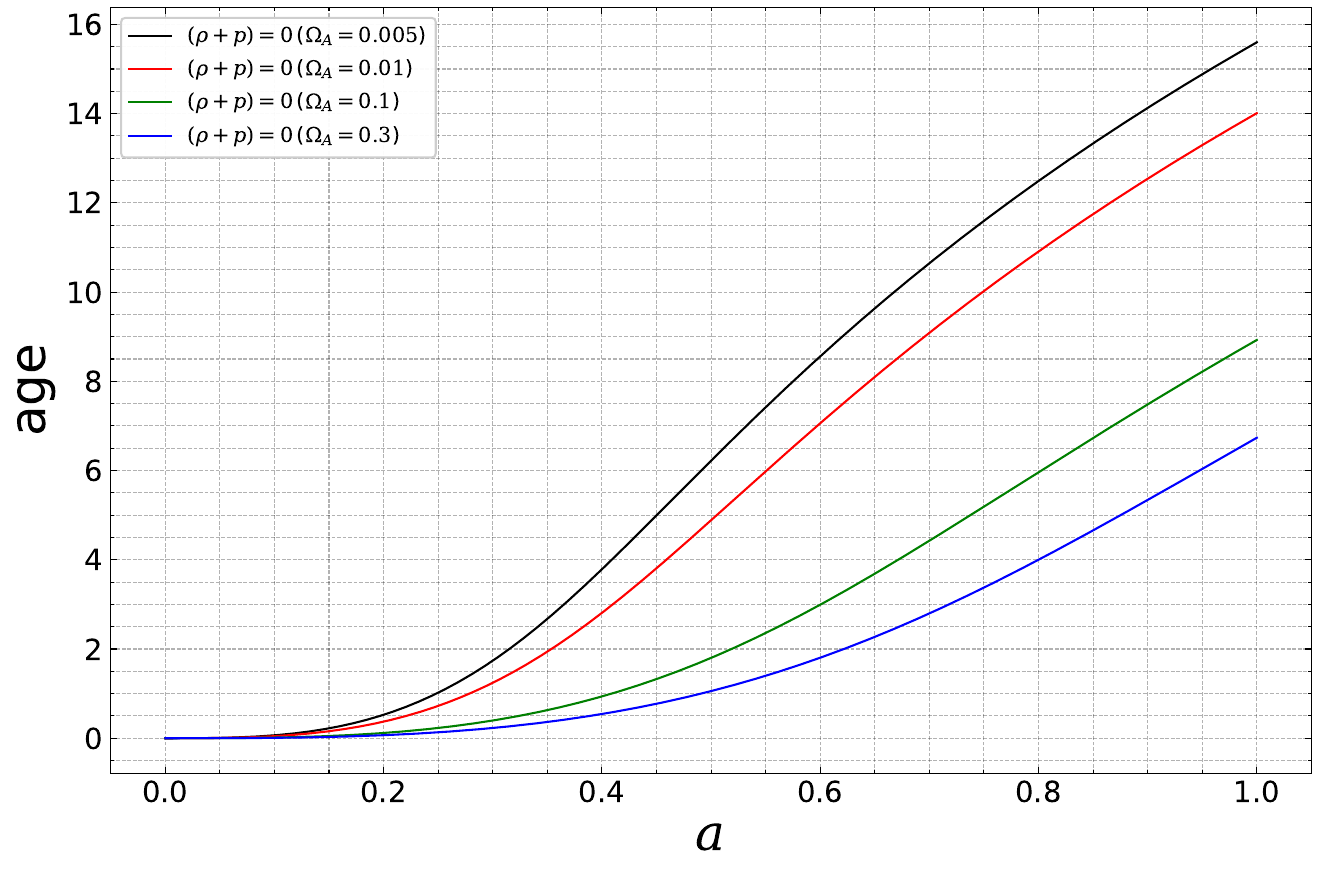}
        \caption{Graph showing the age of the universe for the case $\left(\rho+p\right)=0$. The solid lines represent variations in the parameter $\Omega_{A}$.}
        \label{fig:subfigvacumm}
    \end{subfigure}
    \hfill
    \begin{subfigure}[t]{0.48\textwidth}
        \includegraphics[width=\textwidth]{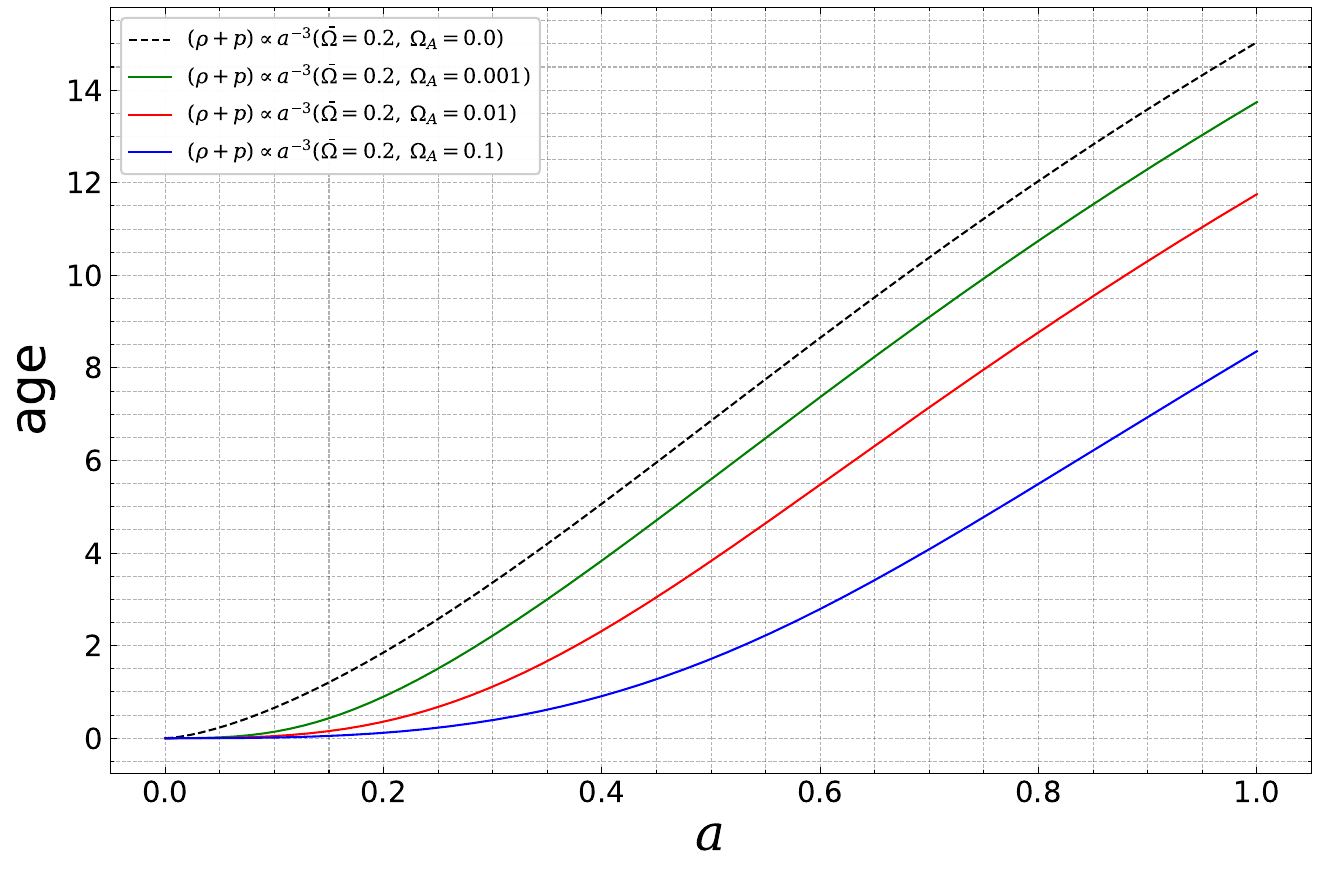}
        \caption{Graph showing the age of the universe for the case  $\left(\rho+p\right)=\bar{\rho}_{0}/a^{-3}$. The solid lines represent variations in the parameter $\Omega_{A}$. The dashed line represent the isotropic limite $\Omega_{A}=0$.}
        \label{fig:subfig2}
   \end{subfigure}
    
     \vspace{0.5cm} 

    \begin{subfigure}[t]{0.48\textwidth}
      \includegraphics[width=\textwidth]{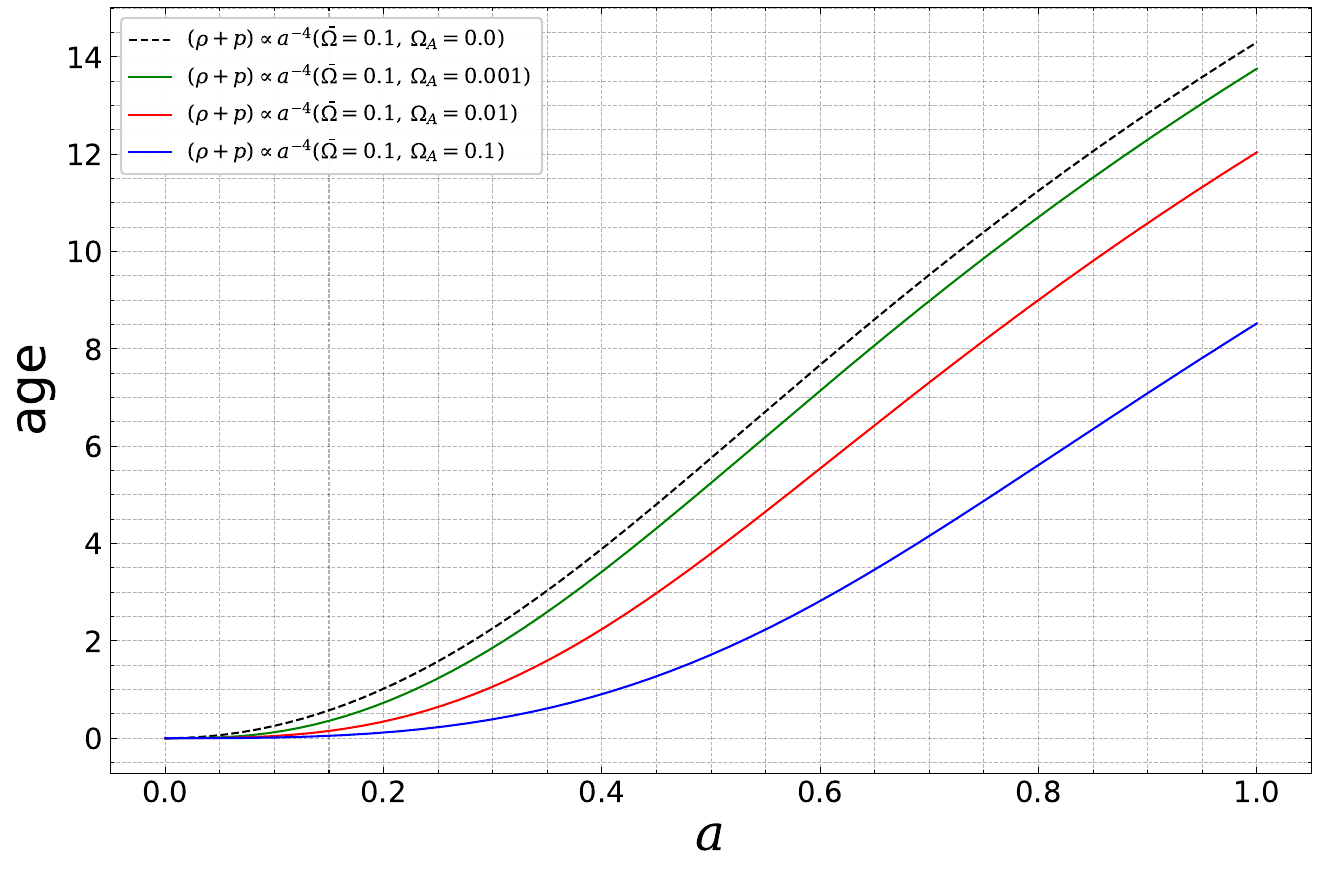}
        \caption{Graph showing the age of the universe for the case  $\left(\rho+p\right)=\bar{\rho}_{0}/a^{-4}$. The solid lines represent variations in the parameter $\Omega_{A}$. The dashed line represent the isotropic limite $\Omega_{A}=0$.}
        \label{fig:subfig3}
    \end{subfigure}
    \hfill
    \begin{subfigure}[t]{0.48\textwidth}
        \includegraphics[width=\textwidth]{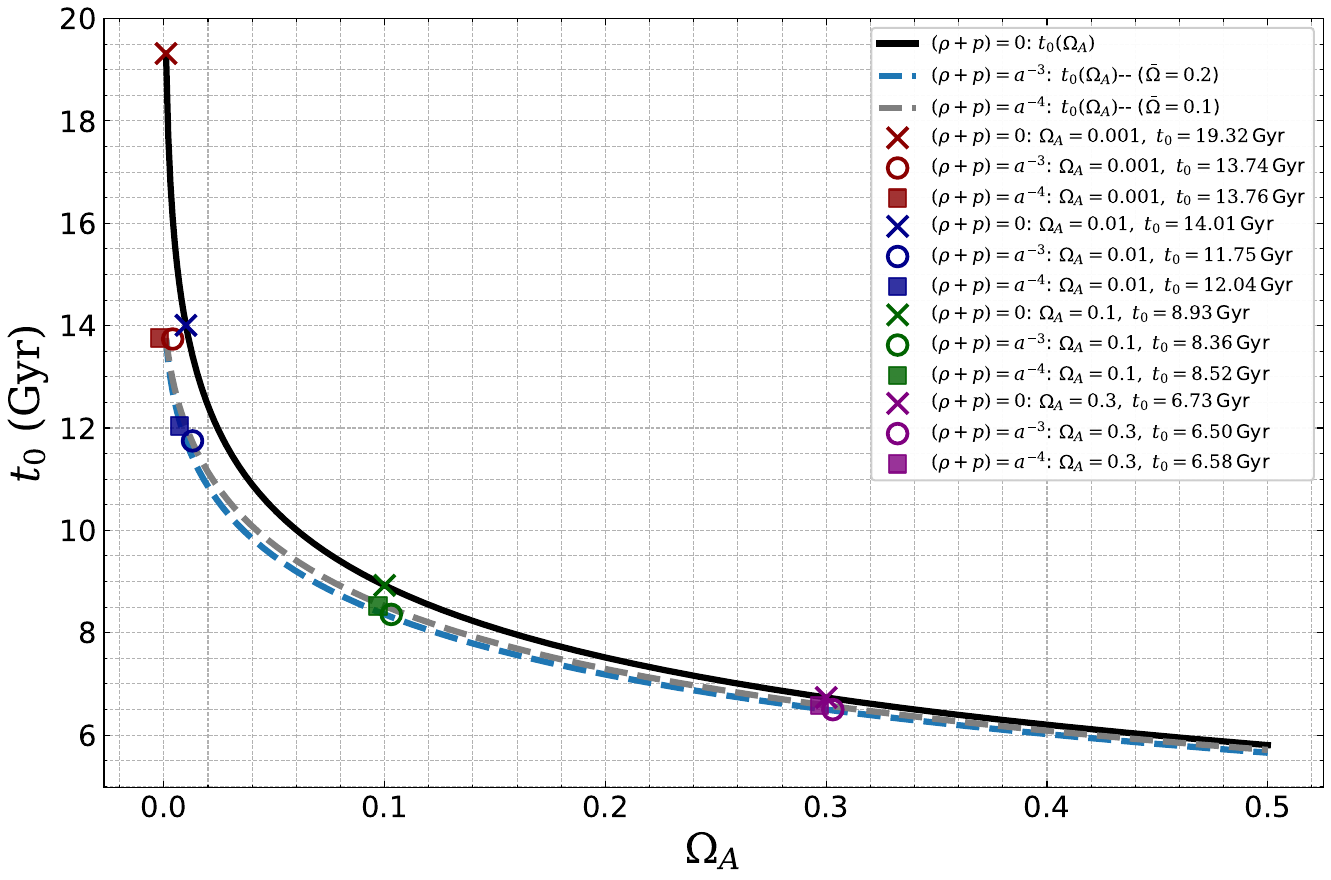}
        \caption{Graph showing the age of the universe as a function of the parameter $\Omega_A$ for the cases $\left(\rho+p\right)=0$, $\left(\rho+p\right)=a^{-3}$ and $\left(\rho+p\right)=a^{-4}$.}
        \label{fig:subfig4}
   \end{subfigure}
\caption{Graph showing the time evolution of the universe for three cases proposed in the combination $\left(\rho+p\right)$.}
    \label{fig:figuraPrincipal}
 \end{figure}

The age of the universe will be given from the integral of the expression
(\ref{eq:43n1}). Using the value of $a\left(t=t_{0}\right)=a_{0}$ in expression (\ref{eq:at}) for the universe
in the present day, we obtain
\begin{align}
t_{0} & =\frac{1}{3\sqrt{\Omega_{U}}H_{0}}\sinh^{-1}\left[\left(\frac{\Omega_{U}}{\Omega_{A}}\right)^{1/2}\right].\label{eq:65age2}
\end{align}
Assuming the fractional energy density parameter $\Omega_{A}=0.01$, $H_{0}=70\,\,km/Mpc \cdot s$ and the identity $\Omega_{U}=1-\Omega_{A}$, we obtain the age of universe $t_{0}\approx 14.0\,\mathrm{Gyr}$. The evolution of the scalar factor in terms of time is represented in figure (\ref{fig:subfigvacumm}), the continuous line represent the vaccum case for the different values of $\Omega_{A}$. We also performed an analysis for this case by sweeping the parameter $\Omega_{A}$ through several values, which can also be seen in the figure (\ref{fig:subfig4}) in the black continuous line. We observed that for increasingly larger values of the parameter $\Omega_{A}$, the age of the universe tends to decrease. For example, at $\Omega_{A}=0.3$, the age of the universe is $t_{0}\approx 7 \,\mathrm{Gyr}$, a difference of half the age compared to the case $\Omega_{A}=0.01$.

In the same way we can obtain the deceleration parameter for this
same model using the expression
\begin{align}
q\left(z\right) & =\frac{H'\left(z\right)}{H\left(z\right)}\left(1+z\right)-1,
\end{align}
were the line represents the derivative with respect to redshift $z$.
Rewriting the expression for the Hubble expansion rate (\ref{eq:63hubble})
in terms of redshift we obtain the following expression for the deceleration
parameter
\begin{align}
q\left(z\right) & =\frac{3\Omega_{A}\left[\frac{\left(1+z\right)}{\left(1+z_{0}\right)}\right]^{6}}{\left\{ \Omega_{U}+\Omega_{A}\left[\frac{\left(1+z\right)}{\left(1+z_{0}\right)}\right]^{6}\right\} }-1,\label{eq:qz_vacumm}
\end{align}
were $z_{0}$ is value of $z$ at some fixed cosmic time. If we consider $a_{0}=1\,\left(z_{0}=0\right)$ for the universe today, the deceleration parameter today is given by
\begin{align}
    q_{0}&=\frac{3\Omega_{A}}{\left[\Omega_{U}+\Omega_{A}\right]}-1.
\end{align}
Using identity $\Omega_{U}=1-\Omega_A$ the expression for the deceleration parameter (\ref{eq:qz_vacumm}) becomes
\begin{align}
    q\left(z\right)= \frac{3 \, \Omega_{A} {\left(1+z\right)}^{6}}{\Omega_{A} {\left(1+z\right)}^{6} - \Omega_{A} + 1} - 1
\end{align}
The evolution of the deceleration parameter with respect to redshift for the vacuum case is shown in figure (\ref{fig:placeholder}) solid red line. We can clearly see that an accelerated expansion phase is allowed for this case reaching a deceleration parameter today $q_{0}=-0.97$. To create the graph we set the value of $\Omega_A=0.01$.

In the next section we will analyze the case where we have the combination
$\left(\rho+p\right)=l=\mathrm{cte}$.

\section{Case $\left(\rho+p\right)=l=\mathrm{cte}$.}\label{sec:4}
In this section, we will assume the combination $\left(\rho+p\right)=l$ as the solution. This solution was first analyzed in the context of Unimodular Gravity in the description of a homogeneous and isotropic cosmological model in \cite{Alvarez_2021}, obtaining a solution for the scale length that has an asymptotically an attractor corresponding to an acelerate universe. Unlike what was presented in \cite{Alvarez_2021}, we will assume the combination $\left(\rho+p\right)=l$ in the anisotropic structure in $\mathrm{NUG}$.
In this case, the equations that dictate the cosmological dynamics of
the anisotropic model in $\mathrm{NUG}$, eqs. (\ref{eq:35}) and (\ref{eq:36}),
are as follows
\begin{align}
\frac{\ddot{a}}{a}-\frac{\dot{a}^{2}}{a^{2}}+\frac{1}{4}\left[3X_{1}^{2}+X_{2}^{2}\right]\left(\frac{a}{a_{0}}\right)^{-6} & =-4\pi Gl\label{eq:63n}\\
\frac{\dddot{a}}{a}+\frac{\ddot{a}\dot{a}}{a^{2}}-2\frac{\dot{a}^{3}}{a^{3}}-\frac{1}{2}\left\{ 3X_{1}^{2}+X_{2}^{2}\right\} \left(\frac{a}{a_{0}}\right)^{-7}\left(\frac{\dot{a}}{a_{0}}\right) & =-16\pi G\frac{\dot{a}}{a}l.\label{eq:64n}
\end{align}
Let's rewrite the above expressions considering the substitution $\left(v\rightarrow\frac{a}{a_{0}}\right)$,
getting
\begin{align}
\frac{\ddot{v}}{v}-\frac{\dot{v}^{2}}{v^{2}}+\frac{\left[3X_{1}^{2}+X_{2}^{2}\right]}{4}v^{-6}+4\pi Gl & =0\label{eq:65n}\\
\frac{\dddot{v}}{v}+\frac{\ddot{v}\dot{v}}{v^{2}}-2\frac{\dot{v}^{3}}{v^{3}}-\frac{\left[3X_{1}^{2}+X_{2}^{2}\right]}{2}v^{-7}\dot{v}+16\pi Gl\frac{\dot{v}}{v} & =0.\label{eq:66n}
\end{align}
Now, inspecting expression (\ref{eq:66n}) we extract the following
equation
\begin{align}
\frac{\ddot{v}}{v}+\frac{\dot{v}^{2}}{v^{2}}+\frac{\left[3X_{1}^{2}+X_{2}^{2}\right]}{12}v^{-6}+16\pi G\ln\left(v\right)^{l} & =\frac{2}{3}\Lambda_{U},\label{eq:67n}
\end{align}
where $\Lambda_{U}$ is a integration constant, associated with cosmological
constant. 
Therefore, the complete set of equations that will describe the cosmological
dynamics of an anisotropic model in $\mathrm{NUG}$ for a fluid that behaves
as constant $\left(\rho+p\right)=l$ is given by equations
(\ref{eq:65n}) and (\ref{eq:67n}). The combination of these equations
results in the following expression
\begin{align}
\left(\frac{\dot{a}}{a}\right)^{2}&=\frac{\Lambda_{U}}{3}-8\pi Gl\left(\ln\left(\frac{a}{a_{0}}\right)-\frac{1}{4}\right)+\frac{\left[3X_{1}^{2}+X_{2}^{2}\right]}{12}\left(\frac{a}{a_{0}}\right)^{-6},\label{eq:68n}
\end{align}
where we have reviewed the change of variable $v\rightarrow\frac{a}{a_{0}}$.

We begin by rewriting the expression (\ref{eq:68n}) in terms of Hubble parameter and in its version written in terms of the fractional energy
density parameters, i.e.
\begin{align}
    H^{2}	=H_{0}^{2}\left\{\Omega_{U}-3\Omega_{l}\ln\left(\frac{a}{a_{0}}\right)+\frac{3}{4}\Omega_{l}+\Omega_{A}\left(\frac{a}{a_{0}}\right)^{-6}\right\},\label{eq:65n1} 
\end{align}
where we have defined the following fractional energy density parameters: $\Omega_{U}\equiv\frac{\Lambda_{U}}{3H_{0}^{2}}$, $\Omega_{l}\equiv\frac{8\pi Gl}{3H_{0}^{2}}$ and $\Omega_{A}\equiv\frac{\left[3X_{1}^{2}+X_{2}^{2}\right]}{12H_{0}^{2}}$. The normalyzed Hubble parameter takes the form
\begin{align}
     E\left(a\right)=\left\{-3\Omega_{l}\ln\left(\frac{a}{a_{0}}\right)+\Omega_{A}\left(\frac{a}{a_{0}}\right)^{-6}+1-\Omega_{A}\right\}^{1/2},\label{eq:Ea} 
\end{align}
\begin{figure}
    \centering
    \includegraphics[width=0.5\linewidth]{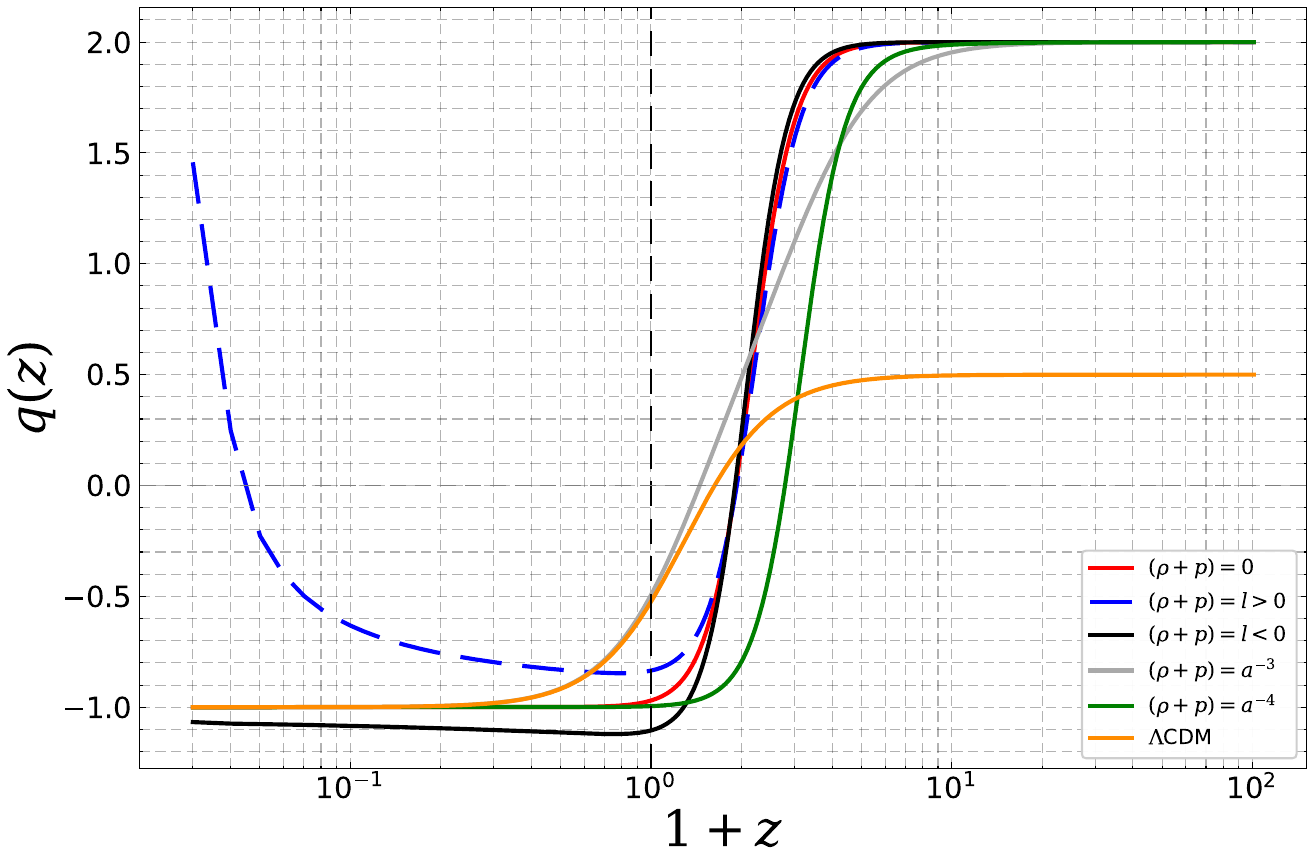}
    \caption{Graph of the evolution of the deceleration parameter $q\left(z\right)$ analyzed in the anisotropic model in NUG plus $\Lambda \mathrm{CDM}$ model. In both graphs, the horizontal axis represented by $\left(1+z\right)$ is on the logarithmic scale. The dashed black line in the graphs above represents the redshift today $z=0$. For the case $\left(\rho+p\right)=0$, represented by the solid red line. To create the graph we set $\Omega_{A}=0.01$. For the case $\left(\rho+p\right)=l$ we plot two cases: $\Omega_{l}>0$ (represented by the dashed blue line) and $\Omega_{l}<0$ (represented by the solid black line). To create the graph we set $\Omega_{A}=0.001$, $\Omega_{l}=\pm 0.09$. For the case $\left(\rho+p\right)=\bar{\rho}_{0}/a^{-3}$, represented by the solid gray line. To create the graph we set $\Omega_{A}=0.01$, $\bar{\Omega}=0.317$. For the case $\left(\rho+p\right)=\bar{\rho}_{0}/a^{-4}$, represented by the solid green line. To create the graph we set $\Omega_{A}=0.001$, $\bar{\Omega}=0.0009$. For the $\Lambda \mathrm{CDM}$ model, represented by solid orange line, we use $\Omega_{m}=0.317$.}
    \label{fig:placeholder}
\end{figure}
where we use the identity $\Omega_{U}=1-\Omega_{A}-\frac{3}{4}\Omega_{l}$. Analyzing the expression (\ref{eq:Ea}), constant enthalpy density leads to a logarithmic factor in the Hubble expansion rate. Therefore, the cosmological dynamics for the asymptotic solutions is strictly dependent on the constant value for the enthalpy density, i.e., $\Omega_{l}>0$ or $\Omega_{l}<0$.

For the case $\Omega_{l}>0$: for small values of the scale factor $a\ll a_{0}$ in (\ref{eq:Ea}) the anisotropic term $\Omega_{A}a^{-6}$ predominates over the others and as the universe expands it decays, recovering isotropy. While for values of the scale factor much larger than today $a\gg a_{0}$ the logarithmic term dominates and becomes increasingly negative (remembering that $\Omega_{l}>0$). Therefore, in this case, $E^{2}\left(a\right)$ will reach negative values in the future, that is, there exists a value $a_{max}=a_{0}\exp\left(\frac{1-\Omega_{A}}{3\Omega_{l}}\right)$ such that $E^{2}\left(a_{max}\right)=0$ and then reaches $E^{2}\left(a\right)<0$ leading to a recollapse. For example, for the values $\Omega_{l}=+0.09$ and $\Omega_{A}=0.001$ assigned in this work $a_{max}\approx 40 a_{0}$, this means that the dynamics of the universe works well (in the sense of accelerated expansion) up to this point, after which the universe enters into recollapse.

For the case $\Omega_{l}<0$: small values in the scale factor $a\ll a_{0}$ in (\ref{eq:Ea}) the anisotropic term is also predominant and we have no significant change. However, for values in the scale factor much larger than today $a\gg a_{0}$ the logarithmic term will dominate and, although, it becomes increasingly negative the choice $\Omega_{l}<0$ causes $E^{2}\left(a\right)>0$ driving a ghost-like super-acceleration.

We can now write the deceleration parameter $q\left(z\right)$ through the expression (\ref{eq:65n1}), obtaining
\begin{align}
    q\left(z\right)=\frac{3 \, {\left(2 \, \Omega_{A} {\left(1+z\right)}^{5} + \frac{\Omega_{l}}{1 + z}\right)} {\left(1+z\right)}}{2 \, {\left(\Omega_{A} {\left(1+z\right)}^{6} -3\Omega_{l} \ln\left({\frac{1}{1+z}}\right) - \Omega_{A} + 1\right)}} - 1, \label{eq:qzlcte}
\end{align}
The evolution of the deceleration parameter for the case $\left(\rho+p\right)=l$ is shown in the figure (\ref{fig:placeholder}). When we have $\Omega_{l}>0$ we represent it as blue dashed lines, while for $\Omega_{l}<0$ we represent it as a black solid line. Here we can see, that for the case $\Omega_{l}>0$ the model allows for an accelerated expansion phase reaching a deceleration parameter today $q_{0}=-0.83$ for $\Omega_{A}=0.01$ and $\Omega_{l}=0.09$. However, as already discussed and which can also be seen in the graph of $q(z)$ (\ref{fig:placeholder}), in the future the model transitions from an accelerated phase to a decelerated phase, recolapsing at $z\approx-1$.

On the other hand, the case $\Omega_{l}<0$ the model allows for an accelerated expansion phase reaching a deceleration parameter today $q_{0}=-1.10$ for $\Omega_{A}=0.01$ and $\Omega_{l}=-0.09$. However, unlike the previous case, this model is characterized by not having a recollapse in the future, that is, the universe in the future remains accelerated as can be seen in the solid black line in (\ref{fig:placeholder}).

Regarding the isotropic limit, it will occur when $X_{1}$ and $X_{2}$ in (\ref{eq:34}) are identically zero and consequently when $\Omega_{A}=0$ in (\ref{eq:65n1}). Therefore, in the isotropic limit, the Hubble expansion rate will be
\begin{align}
   H^{2}	=H_{0}^{2}\left\{\Omega_{U}-3\Omega_{l}\ln\left(\frac{a}{a_{0}}\right)+\frac{3}{4}\Omega_{l}\right\}.\label{eq:IsoH} \end{align}
The solution of the expression (\ref{eq:IsoH}) for the scale factor produces
\begin{align}
    a \left(t\right)=\exp{\left(H_{0}\Delta t-\frac{3}{4}\Omega_{l}H_{0}^{2}\Delta t^{2}\right)},\label{eq:aiso}
\end{align}
where we use $a(t=t_0)=a_{0}=1$, the identity $\Omega_{U}+\frac{3}{4}\Omega_{l}=1$ and defined $\Delta t\equiv t-t_{0}$. Hence, if $\Omega_{l} > 0$, the universe begins from a singular state $(a = 0)$ at $\Delta t\rightarrow -\infty$,
reaching a maximum value and recolapsing to a singular state at $\Delta t\rightarrow + \infty$. If $\Omega_{l} < 0$ the universe evolves from an infinity value to another infinity value passing
by a minimum. This is a non-singular bouncing universe.
The question is the meaning of a negative value for $\Omega_{l}$. However, this can be connected with a violation of the null energy condition, that is, a phantom fluid.

The solution for the scale factor in the isotropic limit (\ref{eq:aiso}), is different from that obtained by the authors in \cite{Alvarez_2021}. In this reference the authors imposes the usual conservation of energy and employ a particular time parametrization dictated by the unimodular condition. However, we show that the physics depends strictly on the value of the constant $l$ as presented in \cite{Alvarez_2021}.

The next analisis will be to consider the combination $\bar{\rho}\equiv\left(\rho+p\right)=\bar{\rho}_{0} \left(\frac{a}{a_{0}}\right)^{-n}$,
with $n=3$.

\section{Case $\left(\rho+p\right)\propto \left(\frac{a}{a_{0}}\right)^{-3}$.}\label{sec:5}

For this case we must note that the expressions (\ref{eq:35}) and (\ref{eq:36}) become
\begin{align}
    \frac{\ddot{a}}{a}-\frac{\dot{a}^{2}}{a^{2}}+\frac{1}{4}\left[3X_{1}^{2}+X_{2}^{2}\right]\left(\frac{a}{a_{0}}\right)^{-6}	&=-4\pi G\bar{\rho}_{0}\left(\frac{a}{a_{0}}\right)^{-3}\label{eq:78n}\\
\frac{\dddot{a}}{a}+\frac{\ddot{a}\dot{a}}{a^{2}}-2\frac{\dot{a}^{3}}{a^{3}}-\frac{1}{2}\left\{ 3X_{1}^{2}+X_{2}^{2}\right\} \left(\frac{a}{a_{0}}\right)^{-7}\left(\frac{\dot{a}}{a_{0}}\right)	&=-4\pi G\bar{\rho}_{0}\left(\frac{a}{a_{0}}\right)^{-4}\left(\frac{\dot{a}}{a_{0}}\right)\label{eq:79n}.
\end{align}
In the same way we can rewrite the expression (\ref{eq:79n})  as follows 
\begin{align}
    \frac{d}{dt}\left[\frac{\ddot{a}}{a}+\frac{\dot{a}^{2}}{a^{2}}++\frac{\left[3X_{1}^{2}+X_{2}^{2}\right]}{12}\left(\frac{a}{a_{0}}\right)^{-6}-\frac{4}{3}\pi G\bar{\rho}_{0}\left(\frac{a}{a_{0}}\right)^{-3}\right]	=0\label{eq:80n}
\end{align}
Which has the following expression as its solution
\begin{align}
\frac{\ddot{a}}{a}+\frac{\dot{a}^{2}}{a^{2}}++\frac{\left[3X_{1}^{2}+X_{2}^{2}\right]}{12}\left(\frac{a}{a_{0}}\right)^{-6}-\frac{4}{3}\pi G\bar{\rho}_{0}\left(\frac{a}{a_{0}}\right)^{-3}	=\frac{2}{3}\Lambda_{U},\label{eq:81n}
\end{align}
where $\Lambda_{U}$ is a constant integration.  Combining the expressions (\ref{eq:78n}) and (\ref{eq:81n}) we obtain a single equation that will be given by
\begin{align}
    \left(\frac{\dot{a}}{a}\right)^{2}	=\frac{\Lambda_{U}}{3}+\frac{8\pi G}{3}\bar{\rho}_{0}\left(\frac{a}{a_{0}}\right)^{-3}+\frac{\left[3X_{1}^{2}+X_{2}^{2}\right]}{12}\left(\frac{a}{a_{0}}\right)^{-6},\label{eq:82n}
\end{align}
or in differential form
\begin{align}
    dt&=\frac{\left(\frac{a}{a_{0}}\right)^{-1}d\left(\frac{a}{a_{0}}\right)}{\left[\frac{\Lambda_{U}}{3}+\frac{8\pi G}{3}\bar{\rho}_{0}\left(\frac{a}{a_{0}}\right)^{-3}+\frac{\left[3X_{1}^{2}+X_{2}^{2}\right]}{12}\left(\frac{a}{a_{0}}\right)^{-6}\right]^{\frac{1}{2}}}.
\end{align}
Adjusting the above expression and integrating both sides, we obtain
\begin{align}
t	=\int\left(\frac{a}{a_{0}}\right)^{-1}\left\{ \frac{\Lambda_{U}}{3}+\frac{8\pi G}{3}\bar{\rho}_{0}\left(\frac{a}{a_{0}}\right)^{-3}+\frac{\left[3X_{1}^{2}+X_{2}^{2}\right]}{12}\left(\frac{a}{a_{0}}\right)^{-6}\right\} ^{-\frac{1}{2}}d\left(\frac{a}{a_{0}}\right).\label{eq:83n}
\end{align}
Once again, the integral on the right hand side of (\ref{eq:83n}) has no analytical solution, a numerical solution is required. We can rewrite the expression (\ref{eq:83n}) in terms of the fractional energy density parameters, i.e.
\begin{align}
   t	=\int\left(\frac{aH_{0}}{a_{0}}\right)^{-1}\left\{ \Omega_{U}+\bar{\Omega}\left(\frac{a}{a_{0}}\right)^{-3}+\Omega_{A}\left(\frac{a}{a_{0}}\right)^{-6}\right\} ^{-\frac{1}{2}}d\left(\frac{a}{a_{0}}\right),\label{eq:84n}
\end{align}
where we have defined the following fractional energy density parameters: $\Omega_{U}\equiv\frac{\Lambda_{U}}{3H_{0}^{2}}$, $\bar{\Omega}\equiv\frac{8\pi G\bar{\rho}_{0}}{3H_{0}^{2}}$ and $\Omega_{A}\equiv\frac{\left[3X_{1}^{2}+X_{2}^{2}\right]}{12H_{0}^{2}}$. We also use a Python code program to find the age of the universe for this case as well as shown in figure (\ref{fig:subfig2}). We assumed an initial value for the scale factor of the order of $10^{-5}$ and plotted the age of the universe in $\mathrm{Gyr}$. We also constructed a graph of the age of the universe as a function of $\Omega_{A}$, shown in the figure (\ref{fig:subfig4}) dashed blue line. To create the graph, we fixed the value of $\bar{\Omega}=0.2$. For the value of $\Omega_{A}=0.01$, we obtain the age of the universe $t_{0}\approx 12\, \mathrm{Gyr}$. In this case, we note the same characteristic as in the vacuum case: larger values of $\Omega_{A}$ decrease the age of the universe.

Rewriting the expression (\ref{eq:81n}) in its Hubble expansion rate version given by
\begin{align}
    H^{2}	=H_{0}^{2}\left\{ \Omega_{U}+\Omega_{A}\left(\frac{a}{a_{0}}\right)^{-6}+\bar{\Omega}\left(\frac{a}{a_{0}}\right)^{-3}\right\}.\label{eq:Hubblematter} 
\end{align}
The deceleration parameter $q(z)$ will be given by the expression
\begin{align}
    q\left(z\right)=\frac{3}{2}\frac{{\left(2 \, \Omega_{A} {\left(z + 1\right)}^{5} + \bar{\Omega} {\left(z + 1\right)}^{2}\right)} {\left(z + 1\right)}}{ \Omega_{A} {\left(z + 1\right)}^{6} + \, \bar{\Omega} {\left(z + 1\right)}^{3} - \, \bar{\Omega} -  \, \Omega_{A} + 1} - 1,\label{eq:qzmatter}
\end{align}
where we use the identity $\Omega_{U}=1-\Omega_{A}-\bar{\Omega}$. The evolution of the deceleration parameter for the case $\left(\rho+p\right)=\bar{\rho}_{0}\left(\frac{a}{a_{0}}\right)^{-3}$ is shown in the figure (\ref{fig:placeholder}) solid gray line. We highlight that this case also allows the universe to undergo an accelerated expansion phase and that it reaches a deceleration parameter today $q_{0}=-0.50$ for $\Omega_{A}=0.01$ and $\bar{\Omega}=0.317$.

Again, the isotropic limit occurs when $X_1$ and $X_2$ are identically zero, and therefore $\Omega_{A}=0$ in (\ref{eq:Hubblematter}). The Hubble expansion rate, in the isotropic limit, will be given by
\begin{align}
    H^{2}	=H_{0}^{2}\left\{ \Omega_{U}+\bar{\Omega}\left(\frac{a}{a_{0}}\right)^{-3}\right\}.\label{eq:Isomatter}
\end{align}
The expression (\ref{eq:Isomatter}) is similar to the cosmological model described by $\mathrm{GR}$, where the cosmic fluid permeating the universe is dust $\left(p=0\right)$ plus a cosmological constant $\left(\Lambda\right)$. However, it is worth highlighting two important points in the expression (\ref{eq:Isomatter}). The term $\Omega_{U}$ in $\mathrm{NUG}$ is associated with a simple integration constant without any (in principle) connection to the vacuum energy density. Furthermore, we recall that the analysis here is done considering that the combination $\left( \rho + p\right)$ has typical dust behavior, meaning that all fluid components in this model will have this behavior.

The next analisis will be to consider the combination $\bar{\rho}\equiv\left(\rho+p\right)=\bar{\rho}_{0} \left(\frac{a}{a_{0}}\right)^{-n}$,
with $n=4$.

\section{Case $\left(\rho+p\right)\propto\left(\frac{a}{a_{0}}\right)^{-4}$.}\label{sec:6}

For this specific case we must note that the right side of the equation
(\ref{eq:36}) is satisfied and equal to zero, that is
\begin{align}
\frac{3}{4}\dot{\bar{\rho}}+3\frac{\dot{a}}{a}\bar{\rho} & =0,\label{eq:71n}
\end{align}
which has its solution exactly the condition we assumed
\begin{align}
\bar{\rho} & =\bar{\rho}_{0}\left(\frac{a}{a_{0}}\right)^{-4},\label{eq:72n}
\end{align}
where we define $\bar{\rho}\equiv\left(\rho+p\right)$. Therefore, the expressions (\ref{eq:35}) and (\ref{eq:36}) take
the form
\begin{align}
\frac{\ddot{a}}{a}-\frac{\dot{a}^{2}}{a^{2}}+\frac{1}{4}\left[3X_{1}^{2}+X_{2}^{2}\right]\left(\frac{a}{a_{0}}\right)^{-6} & =-4\pi G\bar{\rho}_{0}\left(\frac{a}{a_{0}}\right)^{-4}\label{eq:73n}\\
\frac{\dddot{a}}{a}+\frac{\ddot{a}\dot{a}}{a^{2}}-2\frac{\dot{a}^{3}}{a^{3}}-\frac{\left[3X_{1}^{2}+X_{2}^{2}\right]}{2}\left(\frac{a}{a_{0}}\right)^{-7}\left(\frac{\dot{a}}{a_{0}}\right) & =0.\label{eq:74n}
\end{align}
The expression (\ref{eq:74n}) can be rewritten and solved as in the
section $\left(1.1\right)$, that is
\begin{align}
\frac{\ddot{a}}{a}+\frac{\dot{a}^{2}}{a^{2}}+\frac{\left[3X_{1}^{2}+X_{2}^{2}\right]}{12}\left(\frac{a}{a_{0}}\right)^{-6} & =\frac{2}{3}\Lambda_{U},\label{eq:75n}
\end{align}
where $\Lambda_{U}$ is an integration constant. Combining the expressions
(\ref{eq:73n}) and (\ref{eq:75n}) we obtain a single equation that
will be given by
\begin{align}
\left(\frac{\dot{a}}{a}\right)^{2}&=\frac{\Lambda_{U}}{3}+2\pi G\bar{\rho}_{0}\left(\frac{a}{a_{0}}\right)^{-4}+\frac{\left[3X_{1}^{2}+X_{2}^{2}\right]}{12}\left(\frac{a}{a_{0}}\right)^{-6},\label{eq:76n}
\end{align}
or in differential form
\begin{align}
    dt	=&\frac{\left(\frac{a}{a_{0}}\right)^{-1}d\left(\frac{a}{a_{0}}\right)}{\left[\frac{\Lambda_{U}}{3}+2\pi G\bar{\rho}_{0}\left(\frac{a}{a_{0}}\right)^{-4}+\frac{\left[3X_{1}^{2}+X_{2}^{2}\right]}{12}\left(\frac{a}{a_{0}}\right)^{-6}\right]^{\frac{1}{2}}}.
\end{align}
As in the previous case, the age of the universe can be calculated according to the expression
\begin{align}
t	=&\int\left(\frac{a}{a_{0}}\right)^{-1}\left\{ \frac{\Lambda_{U}}{3}+2\pi G\bar{\rho}_{0}\left(\frac{a}{a_{0}}\right)^{-4}+\frac{\left[3X_{1}^{2}+X_{2}^{2}\right]}{12}\left(\frac{a}{a_{0}}\right)^{-6}\right\} ^{-\frac{1}{2}}d\left(\frac{a}{a_{0}}\right).\label{eq:77n}
\end{align}
Once again, the integral on the right hand side of (\ref{eq:77n})
has no analytical solution, a numerical solution is required. We can rewrite the expression (\ref{eq:77n}) in terms of the fractional energy density parameters, i.e. 
\begin{align}
    t	=&\int\left(\frac{aH_{0}}{a_{0}}\right)^{-1}\left\{ \Omega_{U}+\frac{3}{4}\bar{\Omega}\left(\frac{a}{a_{0}}\right)^{-4}+\Omega_{A}\left(\frac{a}{a_{0}}\right)^{-6}\right\} ^{-\frac{1}{2}}d\left(\frac{a}{a_{0}}\right),
\end{align}
where we have defined the following fractional energy density parameters: $\Omega_{U}\equiv\frac{\Lambda_{U}}{3H_{0}^{2}}$, $\bar{\Omega}\equiv\frac{8\pi G\bar{\rho}_{0}}{3H_{0}^{2}}$ and $\Omega_{A}\equiv\frac{\left[3X_{1}^{2}+X_{2}^{2}\right]}{12H_{0}^{2}}$. We also use a Python code program to find the age of the universe for this case as well as shown in figure (\ref{fig:subfig3}). We assumed an initial value for the scale factor of the order of $10^{-5}$ and plotted the age of the universe in $\mathrm{Gyr}$. We also constructed a graph of the age of the universe as a function of $\Omega_{A}$ shown in the figure (\ref{fig:subfig4}) dashed gray line. To create the graph, we fixed the value of $\bar{\Omega}=0.1$. For the value of $\Omega_{A}=0.01$, we obtain the age of the universe $t_{0}\approx 12\, \mathrm{Gyr}$. For this case, we note the same characteristic as the vacuum case and the case $\left(\rho+p\right)=a^{-3}$, larger values of $\Omega_{A}$ decrease the age of the universe.

We begin by rewriting the expression (\ref{eq:76n}) in terms of Hubble parameter and in its version written in terms of the fractional energy
density parameters, i.e.
\begin{align}
   H^{2}	=H_{0}^{2}\left\{ \Omega_{U}+\Omega_{A}\left(\frac{a}{a_{0}}\right)^{-6}+\frac{3}{4}\bar{\Omega}\left(\frac{a}{a_{0}}\right)^{-4}\right\}.\label{eq:Hubbleradiation} 
\end{align}
We can now write the deceleration parameter $q\left(z\right)$ through the expression (\ref{eq:Hubbleradiation}), obtaining
\begin{align}
    q\left(z\right)=\frac{6 \, {\left(2 \, \Omega_{A} {\left(z + 1\right)}^{5} + \bar{\Omega} {\left(z + 1\right)}^{3}\right)} {\left(z + 1\right)}}{4 \, \Omega_{A} {\left(z + 1\right)}^{6} + 3 \, \bar{\Omega} {\left(z + 1\right)}^{4} - 3 \, \bar{\Omega} - 4 \, \Omega_{A} + 4} - 1,\label{eq:qzradiation}
\end{align}
where we use the identity $\Omega_{U}=1-\Omega_{A}-\frac{3}{4}\bar{\Omega}$. The evolution of the deceleration parameter for the case $\left(\rho+p\right)=\bar{\rho}_{0}\left(\frac{a}{a_{0}}\right)^{-4}$ is shown in the figure (\ref{fig:placeholder}) solid green line. We highlight that this case also allows the universe to undergo an accelerated expansion phase and that it reaches a deceleration parameter today $q_{0}=-0.99$ for $\Omega_{A}=0.001$ and $\bar{\Omega}=0.0009$.

Similarly, by taking $X_1$ and $X_2$ identically zero, that is, by taking $\Omega_{A}=0$, we reach the isotropic limit in (\ref{eq:Hubbleradiation}). Therefore, the Hubble expansion rate in this limit will be given by
\begin{align}
    H^{2}	=H_{0}^{2}\left\{ \Omega_{U}+\tilde{\Omega}\left(\frac{a}{a_{0}}\right)^{-4}\right\}, \label{eq:Isoradiation}
\end{align}
where we have defined $\tilde{\Omega}\equiv \frac{3}{4}\bar{\Omega}$. The expression (\ref{eq:Isoradiation}) is the same as that studied in article \cite{Fabris_2022}. It describes the cosmological dynamics of a universe transitioning from an initial radiative phase of decelerated expansion evolving into an accelerated de Sitter-type expansion phase. This dynamics is similar to the radiative model plus cosmological constant described by $\mathrm{GR}$. However, we recall that our analysis starts from the case where we have the combination $\left( \rho +p\right)$ with typical radiation behavior, which means that all fluid components in this model should have this behavior; for more details see \cite{Fabris_2022}.

\section{Conclusions}\label{sec:7}
In summary, we studied an anisotropic Bianchi I-type cosmological model in non-conservative Unimodular Gravity. One of the main striking characteristics in $\mathrm{NUG}$ is the underdetermination of the complete set of equations that describe the cosmological dynamics; that is, to solve it, we need to provide extra information to the set of equations. We saw that studying an anisotropic cosmological model in NUG does not resolve this pathology in its structure. To circumvent this situation, we propose extra information about the combination $\left(\rho+p \right)$.

For the case $\left(\rho+p \right)=0$, the vacuum case, we obtain a cosmological dynamic very close to that described according to $\mathrm{GR}+\Lambda$, as discussed in section (\ref{sec:3}). We find an analytical solution for the scale factor (\ref{eq:at}) and asymptotically the universe experiences a decelerated expansion $(t \rightarrow 0^{+})$ evolving into an accelerated de Sitter-type expansion $(t \rightarrow +\infty)$. For values of $\Omega_{A}=0.01$ and, consequently, $\Omega_{U}=0.99$ we find the age of the universe to be on the order of $\approx 14.0 \, \mathrm{Gyr}$. Kasner's solution is also a solution in $\mathrm{NUG}$, as we demonstrated in subsection (\ref{Kasner}). From Kasner's conditions for the indices $p_{i}$, we obtain a relationship between the constants $X_{1}$, $X_{2}$, and $\Omega_{A}$ that is exactly the same as the one we defined. This fact demonstrates that the parameter $\Omega_{A}$ is connected in the geometry of the Bianchi I-type metric and the field equations in $\mathrm{NUG}$.

We also carried out a completely new analysis which proved to be quite intriguing; the combination $\left(\rho+p \right)=l=cte$, in section (\ref{sec:4}). This mere consideration leads to a logarithmic term appearing in the Hubble expansion rate (\ref{eq:65n1}). This is typical of a non-barotropic cosmic fluid. We find from the analysis of this case that the cosmological dynamics in the future are strictly dependent on the sign of the constant $l$. For $l>0$ $\left(\Omega_{l}>0\right)$ the universe in the future reaches a maximum value and recolapses. Therefore, problems with $E^{2}\left(a\right)<0$ lead to a loss of credibility of the model. On the other hand, the case $l<0$ $\left(\Omega_{l}<0\right)$ demonstrated a more interesting characteristic in the cosmological sense. In the future, the term that has a logarithmic value becomes positive, and the universe finds itself in a super-accelerated state, as highlighted in the figure (\ref{fig:placeholder}). When we recover the isotropy, the characteristics remain for both cases. If we consider $\Omega_{l}<0$, the universe evolves from one infinite value to another infinite value, passing through a minimum, which would be a non-singular and eternal universe. Exploring this solution in other cosmological concepts, such as the description of gravitational waves and the formation of structures in the universe, should be key to a better understanding of $\mathrm{NUG}$. Furthermore, the thermodynamic context can also be explored in an attempt to clarify enthalpy, since the combination $\left(\rho+p \right)$ can be interpreted as the enthalpy density of the system.

For the cases where the combination $\left(\rho+p\right)\propto a^{-3}$ and $\left(\rho+p\right)\propto a^{-4}$ in $\mathrm{NUG}$, we find something very similar to the description of an anisotropic (isotropic) cosmological model according to $\mathrm{GR}+\Lambda$. There are two points that differentiate both approaches; the first is the term $\Omega_{U}$ which appears naturally (in all cases analyzed) as a simple integration constant, and we highlight its important role in cosmological dynamics, similar to $\mathrm{CC}$. Furthermore, a second point is the combination $\left(\rho+p\right)$, which, we emphasize once again, is associated with the enthalpy of the system, and in both cases the chosen behavior is valid regardless of the fluid considered; that is, we are not assuming any equation of state of the type $p_{i}=\omega_{i}\rho_{i}$ in choosing the combination $\left(\rho+p\right)$.

We would like to point out that the choices for the combination $\left(\rho+p\right)$ can also be mapped onto models in $\mathrm{GR}$ with a dynamic cosmological term, which also requires an ansatz to obtain a complete set of equations.

Regarding the anisotropic term in the cosmological equations in NUG, for all cases studied it always appears with a factor $a^{-6}$ similar to the case described according to $\mathrm{GR}$. Now, regarding the fractional energy density parameter $\Omega_A$ we note, that in order to have an age of the universe compatible with the age of globular cluster \cite{Ying_2025, Valcin:2021jcg, Correnti_2016}, it must be around $\Omega_A \approx 10^{-2}$. However, we emphasize that to ensure the value of $\Omega_{A}$, a more complete statistical analysis comparing it with cosmological data should be applied.

\bigskip
\noindent
{\bf Acknowledgments:} We thank CNPq, FAPES, FAPEMIG and CAPES for partial financial support.
 
 \bibliography{bibli1}{}

\end{document}